\documentclass[twocolumn,showpacs,preprintnumbers,amsmath,amssymb]{revtex4}

\usepackage{graphicx}
\usepackage{xcolor}
\definecolor{indigo(dye)}{rgb}{0.0,0.25,0.42}

\begin{document}

\title{Antiferromagnetic Heisenberg Model on the Icosahedron: Influence of Connectivity and the Transition from the Classical to the Quantum Limit}

\author{N. P. Konstantinidis}
\affiliation{Max Planck Institut f\"ur Physik komplexer Systeme, 01187 Dresden, Germany}

\date{\today}

\begin{abstract}
The antiferromagnetic Heisenberg model on the icosahedron, which consists of 20 edge-sharing triangles and belongs to the icosahedral $I_h$ symmetry group, presents unconventional properties at the classical and quantum level. These originate in the frustrated nature of the interactions between the spins. For classical spins the magnetization is discontinuous in a magnetic field \cite{Schroeder05}. In the quantum limit there are non-magnetic excitations within the singlet-triplet gap and the specific heat has a multi-peak structure as function of temperature \cite{NPK05}. Here we examine the importance of the connectivity of the icosahedron for the appearance of the magnetization discontinuity, and also investigate the transition from the classical to the quantum limit. The influence of connectivity on the magnetic properties is revealed by considering the cluster as being made up of a closed strip of a triangular lattice with two additional spins attached. The classical magnetization discontinuity is shown to evolve continuously from the discontinuity effected by these two spins when they are uncoupled to the cluster.
In the second part the transition from the classical to the quantum limit is examined by
focusing on the low energy spectrum taking fully into account the spatial and the spin symmetry of the model in the characterization of the states.
A symmetry analysis of the highly degenerate due to the connectivity lowest energy classical manifold identifies as its direct fingerprint the low energy quantum states for spin magnitude as low as $s=1$, with the latter following a tower of states behavior which relates to the icosahedron having a structure reminiscent of a depleted triangular lattice. The classical character of the AHM for small $s$ is also detected on the ground state energy and correlation functions. On the other hand the classical magnetization discontinuity in a field eventually disappears for small $s$, after a weak reentrant behavior.
\end{abstract}


\maketitle

\section{Introduction}
\label{sec:introduction}
The antiferromagnetic (AFM) Heisenberg model (AHM) is a prototype for the investigation of strongly correlated electronic behavior \cite{Auerbach98,Fazekas99}. Its properties depend on the interplay of dimensionality, strength of quantum fluctuations and frustration generated by the particular connectivity of a lattice or cluster. It has been extensively studied on different types of lattices with the hope that frustration will lead to the long sought after spin liquid phase \cite{Anderson73}. It has also been considered on finite clusters to model their magnetic behavior, especially since recent advances in synthetic chemistry have led to the production of molecules with important magnetic properties that form the class of molecular nanomagnets and could provide the building blocks for quantum computers and memory devices \cite{Gatteschi06}. Another route for the fabrication of small entities with well controlled magnetic properties is provided by scanning tunneling microscopy, where the AHM is also of utmost importance \cite{NPK13}. In all the above cases the analysis of collective magnetic behavior is of fundamental importance. This necessitates the determination of magnetic properties for various clusters with different connectivities as function of the spin magnitude of the magnetic units $s$, which can be done firstly within the framework of the AHM in order to search for correlations between structure and magnetic behavior.

A direct way to control magnetic properties is through the application of a magnetic field. A Hamiltonian isotropic in spin space switches at the quantum level between neighboring total spin $S$ sectors where $\Delta S=1$ at appropriate values of the field. In frustrated clusters there can be unexpected discontinuities that skip certain magnetic sectors and have $\Delta S>1$ \cite{Schnack10}. It is known that for structures with the connectivity of the fullerenes the magnetization in a field can be discontinuous at the classical level \cite{Coffey92}. Multiple discontintinuities were also found at the classical and extreme quantum limit for the smallest member of the fullerene family, the dodecahedron \cite{NPK05}, as well as other fullerene clusters that share the icosahedral $I_h$ point group symmetry with the dodecahedron \cite{NPK07}. For small fullerene clusters of other symmetry only pronounced magnetization plateaus were found for $s=1/2$ \cite{NPK09}. It was also discovered that the icosahedron, the smallest cluster with $I_h$ symmetry which is nevertheless not a member of the fullerene family, has a magnetization discontinuity at the classical level which persists for lower values of $s$ \cite{Schroeder05}. Additionally for small $s$ there are non-magnetic excitations within the singlet-triplet gap and the specific heat has a multi-peak structure as function of temperature for both the dodecahedron and the icosahedron \cite{NPK05}. These are indications of a strong correlation between spatial symmetry and magnetic behavior as the unit that causes frustration is different in the two cases, the pentagon in the fullerenes and the triangle in the icosahedron.

Further insight on the appearance of magnetization discontinuities can be gained by considering the structure of the molecules more closely. The icosahedron comprises of 20 edge-sharing triangles, and each of its vertices is five-fold coordinated. Unlike the cuboctahedron and icosidodecahedron for example, which are finite periodic realizations of the Kagom\'e lattice on a sphere \cite{Rousochatzakis08}, there is no direct correspondence between the icosahedron and a periodic lattice, since the latter can not have a five-fold symmetry axis. On the other hand in the triangular lattice each vertex has six nearest-neighbors, so the icosahedron is reminiscent of a closed finite version of the triangular lattice where each vertex is feeling the effect of a single defect, which is suggestive of depleted triangular lattices \cite{Betts95,Schulenburg00,Keene11}. Similarly, the dodecahedron does not correspond to a periodic lattice due to having five-fold symmetry axes \cite{NPK05} (the two dimensional plane can only be tiled using irregular pentagons, as is the case of the Cairo pentagonal lattice that has discontinuities in the magnetization if two different AFM interactions originating in the two kinds of inequivalent sites of the lattice are considered \cite{Nakano14}). For fullerene clusters in general the concept of defects is also important, as the 12 pentagons they include appear so among an arbitrary number of hexagons \cite{Fowler95,Coffey92,NPK07}. The presence of magnetization discontinuities in the icosahedron and the fullerenes appears then to be closely connected with the presence of defects of one form or another. It must also be noted that a single vacancy in a magnetic lattice generates a magnetization discontinuity in the limit of vanishing field \cite{Wollny12}.

In this paper first the importance of the connectivity of the icosahedron for its magnetic properties is investigated. To this end, the cluster is broken up in two parts. The first part is a closed strip of edge-sharing triangles which can also be viewed as a closed chain with equal nearest and next-nearest neighbor interactions, while the second part has two spins which are attached to the top and the bottom of the triangle strip respectively. The interaction of the two extra spins with their closest neighbors on the icosahedron is taken to increase progressively from 0 to 1, which can be viewed as if the two extra spins are slowly brought closer to the rest. When this interaction is zero the magnetic response is determined by the one of the triangle strip, which classically has a constant susceptibility, and the one of the two extra spins, which immediately respond to a magnetic field and therefore generate a magnetization discontinuity at zero field for the whole system. Increasing the coupling of the extra spins with the rest the magnetization discontinuity evolves continuously and survives up to the point where all interactions among the twelve spins are the same. It is then concluded that the special connectivity of the icosahedron allows for a magnetization discontinuity within the framework of the AHM even though there is no spin or spatial anisotropy. It is also shown that close to the spatial isotropic limit another discontinuity appears.

In the second part of the paper the AHM in the spatially isotropic case is considered and the influence of the triangular structure of the icosahedron on the low energy spectrum for arbitrary $S$ is investigated. The icosahedron consists of edge-sharing triangles with five-fold coordinated vertices, and the triangular lattice without or with defects has been shown to possess towers of states in its low energy structure that originate in the large spatial degeneracy of the classical ground states and point to a long range ordered ground state \cite{Bernu94,Schulenburg00}. Here it is examined with a symmetry analysis for how small $s$ the classical lowest energy configurations leave their fingerprint in the low energy spectrum. The classical lowest energy configuration is highly degenerate both below and above the magnetization discontinuity due to frustration, and a symmetry analysis \cite{Rousochatzakis08,Rousochatzakis12} classifies the configurations in these two degenerate manifolds according to either $S$ or the total spin along the $z$ axis $S^z$ and the irreducible representations of $I_h$. In the case of finite $s$ spatial and spin symmetries of the AHM characterize its eigenstates in a similar fashion. More precisely, the low energy spectrum is calculated for $s$ ranging from 1/2 to 5/2.
For $S$ below and above the magnetization discontinuity there is a respective set of eigenstates well separated from higher excited states with a symmetry structure identical with the one of the classical lowest energy manifold, showing that the lowest states for finite $s$ originate directly in the classical degenerate configurations. This Anderson tower of states structure remains invariant for $s \gtrsim 1$ and for an infinite lattice would express SU(2) symmetry breaking. The icosahedron is however finite but its structure is reminiscent of depleted triangular lattices,
consequently a classical description of the icosahedron is a good approximation down to small $s$. A similar conclusion is drawn from the ground state energy and correlation functions, which show that starting from the classical ground state an expansion in a few orders of $\frac{1}{s}$ will produce accurate estimates down to $s=\frac{1}{2}$.

From the lowest energy in every $S$ sector it is possible to calculate the magnetization in a field as function of $s$. It is found that the classical discontinuity which first appears for $s=4$ \cite{Schroeder05} unexpectedly disappears for $s=9/2$, only to reappear for $s=5$ and $11/2$. The symmetry pattern of the lowest eigenstates as function of $S$ on the two sides of the discontinuity is different and follows its corresponding classical counterpart.

The plan of this paper is as follows: in Sec. \ref{sec:modelandclassicallowestenergyconfiguration} the model is presented along with the methods of calculation, and the classical result is summarized. In Sec. \ref{sec:connectivityandmagneticproperties} the relationship between the connectivity of the icosahedron and its magnetic response is examined. Sec. \ref{sec:lowenergyspectrum} establishes the relationship between the quantum low energy spectrum and the classical lowest energy configurations, while in Sec. \ref{sec:magnetizationinafield} the magnetization process in a field is investigated. Finally Sec. \ref{sec:conclusions} presents the conclusions.

\section{Model and Classical Lowest Energy Configuration}
\label{sec:modelandclassicallowestenergyconfiguration}

\begin{figure}
\includegraphics[width=3.5in,height=2.5in]{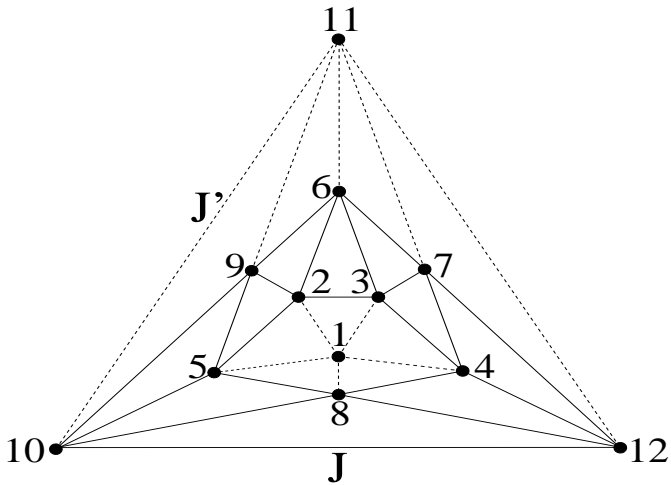}
\vspace{0pt}
\caption{A projection of the icosahedron on a plane. The circles are spins of magnitude $s$ and each interacts with its five nearest neighbors as shown by the connecting lines. The solid and dashed lines indicate interactions of strength $J$ and $J'$ respectively. At the spatially isotropic limit $J'=J$.
}
\label{fig:icosahedronclusterconnectivity}
\end{figure}

A projection of the icosahedron on a plane is shown in Fig. \ref{fig:icosahedronclusterconnectivity}. It consists of 20 triangles and all $N=12$ vertices are equivalent and five-fold coordinated.
The Hamiltonian of the AHM is
\begin{equation}
H = J \sum_{<ij>} \textrm{} [ \frac{1}{2} (s_i^+ s_j^- + s_i^- s_j^+) + s_i^z s_j^z ] - h \sum_{i=1}^N s_i^z
\label{eqn:Hamiltonian}
\end{equation}
$J$ is the strength of the exchange interaction, which is positive and taken to be 1 from now on, defining the unit of energy. $<i,j>$ indicates that interactions are limited to nearest neighbor spins, which have magnitude $s$. The magnetic field $\vec{h}$ is taken along the $\hat{z}$ direction. In Hamiltonian (\ref{eqn:Hamiltonian}) the exhange energy competes with the magnetic energy, with the frustrated connectivity of the icosahedron playing an important role. For finite $s$ as the magnetic field increases $S^z$ increases stepwise, until it saturates. Hamiltonian (\ref{eqn:Hamiltonian}) is block diagonalized according to the point symmetry group $I_h$, and also the spin inversion symmetry group when $S^z=0$ \cite{NPK05}. When the blocks of the Hamiltonian are small enough full diagonalization is possible and the energy spectrum is fully available. Up to now the full energy spectrum has been calculated within the $I_h$ group for $s=1/2$ and 1 \cite{NPK05} and within the $D_2$ subgroup of $I_h$ for $s=3/2$ \cite{Schnalle10}. For large $s$ only Lanczos diagonalization is possible and gives at least the lowest energy level for each $S^z$ in order to produce the magnetization response in the field. Comparison of the energies in different $S^z$ sectors allows characterization of the states also according to $S$, with each $S$ state being $2S+1$ times degenerate. In the classical limit $s \to \infty$ classical minimization is performed \cite{Coffey92,NPK05,NPK07}.

\begin{figure}
\includegraphics[width=3.5in,height=2.5in]{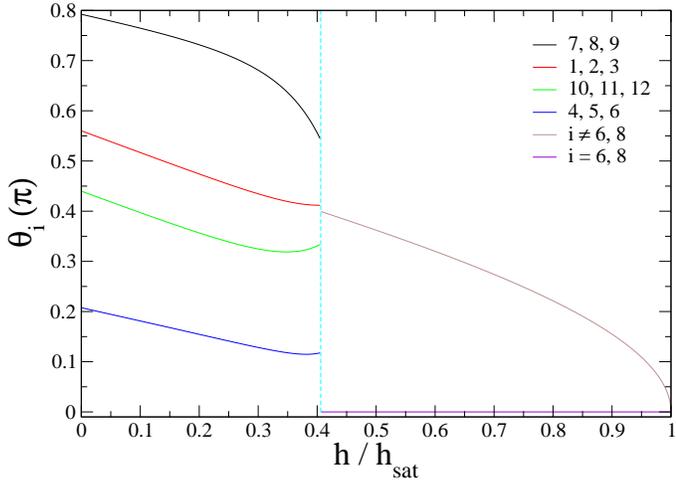}
\vspace{0pt}
\caption{Polar angles $\theta_i(\pi)$ as a function of the magnetic field $h$ over its saturation value $h_{sat}$ in the lowest energy configuration of the icosahedron. The dashed line shows the location of the discontinuity field $\frac{h_{disc}}{h_{sat}}=0.40603$. For the higher field phase $\theta_6=\theta_8=0$ and for the rest of the spins $\theta_i=\textrm{arccos}\frac{\textrm{h}-1}{4+\sqrt{5}}$. The configurations below and above the discontinuity field are degenerate, as the polar angles can be distributed in different ways among the spins to achieve the lowest energy. One possible distribution (coloring) for above and one for below the discontinuity is shown here.
}
\label{fig:icosahedronangles}
\end{figure}

The triangle, the basic unit comprising the icosahedron, supports frustrated interactions between spins sitting on its vertices. When the 20 triangles are brought together to form the icosahedron by sharing edges frustration increases, and the classical nearest-neighbor lowest energy in the absence of a field reduces from the value of $-\frac{1}{2}$ for an isolated triangle to $-\frac{\sqrt{5}}{5}$ \cite{Schmidt03}. The classical lowest energy configuration of Hamiltonian (\ref{eqn:Hamiltonian}) for $h \neq 0$ was found in \cite{Schroeder05}. For lower fields the spins assume four different polar angle values (Fig. \ref{fig:icosahedronangles}), while the azimuthal angles assume six different equidistant values. Starting at $h_{disc}=0.40603 h_{sat}$ (with $h_{sat}=5+\sqrt{5}$) a different configuration minimizes the energy until saturation. Two spins are aligned with the field and the rest share a common polar angle, while their azimuthal angles have ten different equidistant values. The change from the lower to the higher field configuration is accompanied by a magnetization discontinuity.

In the high field configuration an analytic calculation is feasible under the constraint that there is only one angle varying with the field. The common polar angle of the ten spins is $\textrm{arccos}\frac{\textrm{h}-1}{4+\sqrt{5}}$, and the total magnetization $2\frac{5\textrm{h}+\sqrt{5}-1}{4+\sqrt{5}}$. Looking at Fig. \ref{fig:icosahedronangles}, as the discontinuity is approached from below the polar angles counterintuitively increase weakly with the field, except from the one of spins 7, 8, 9. This non-monotonic dependence has also been found for the AHM in open chains and fullerene clusters with $I_h$ symmetry \cite{Machens13,NPK14,NPK05,NPK07}.

\section{Connectivity and Magnetic Response}
\label{sec:connectivityandmagneticproperties}
To investigate the role of connectivity in the magnetization of the icosahedron a spatial decomposition of the cluster in two parts is chosen. This decomposition groups two vertices diametrically opposite to one another, for example vertices 1 and 11 in Fig. \ref{fig:icosahedronclusterconnectivity}, and has the rest forming a structure which is part of a triangular lattice but folds back to itself, having periodic boundary conditions. They can also be viewed as a closed chain with equal nearest and next-nearest neighbor interactions. If the coupling within the triangle strip is $J$ while its spins couple to spins 1 and 11 with exchange strength $J'$ the corresponding Hamiltonian, which is spatially anisotropic when $J' \neq J$, is:
\begin{eqnarray}
H & = & J \sum_{<ij>^,i,j \neq 1,11} \vec{s}_i \cdot \vec{s}_j + ( J' \vec{s}_1 -\vec{h} ) \cdot ( \vec{s}_2 + \vec{s}_3 + \vec{s}_4 + \nonumber \\ & & \vec{s}_5 + \vec{s}_8 ) + ( J' \vec{s}_{11} -\vec{h} ) \cdot ( \vec{s}_6 + \vec{s}_7 + \vec{s}_9 + \vec{s}_{10} + \vec{s}_{12} ) \nonumber \\ & & - \vec{h} \cdot ( \vec{s}_1 + \vec{s}_{11} )
\label{eqn:Hamiltoniantwoparts}
\end{eqnarray}
$J$ will be taken to be 1 from now on. In the absence of a field the angle between nearest neighbors in the triangle strip is $\frac{3\pi}{5}$ or $\frac{4\pi}{5}$. Its magnetic response has constant susceptibility all the way to saturation, with the spins turning towards the field direction having the same polar angle. On the other hand the two isolated spins saturate immediately in a field, therefore the magnetization of the combined system has a discontinuity in zero field when $J'=0$. Consequently an infinitesimal field leads to the high field configuration of the spatially isotropic case as described in Sec. \ref{sec:modelandclassicallowestenergyconfiguration}, with the two weakly coupled spins being the ones aligned with the field. The influence of a finite $J'$ is to effectively reduce the magnetic field on the spins of the triangle strip as seen from the second and third terms of Eq. (\ref{eqn:Hamiltoniantwoparts}), and this effective field does not necessarily lie along the magnetic field axis.

\begin{figure}
\includegraphics[width=3.5in,height=2.5in]{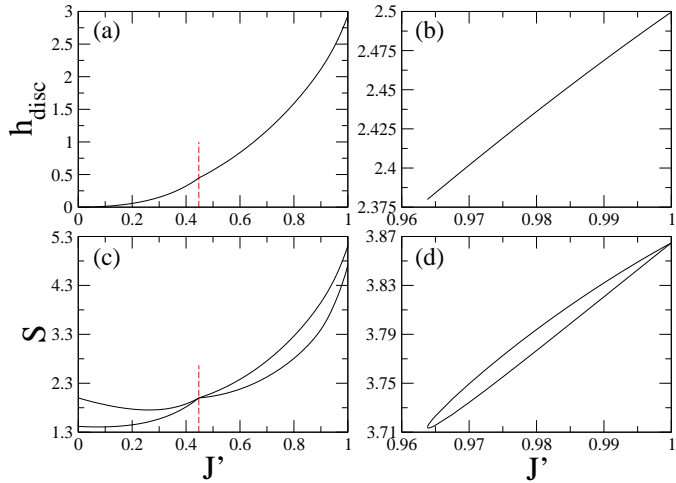}
\vspace{0pt}
\caption{(a) Magnetic field $h_{disc}$ where the magnetization discontinuity takes place and (c) magnetization $S$ just below and just above the discontinuity as functions of $J'$. The red long dashed line at $J'=0.44722$ shows where the discontinuity is one of the susceptibility. (b) and (d) give the corresponding values for the discontinuity that appears for larger $J'$.
}
\label{fig:transitionfield}
\end{figure}

\begin{figure}
\includegraphics[width=3.5in,height=2.5in]{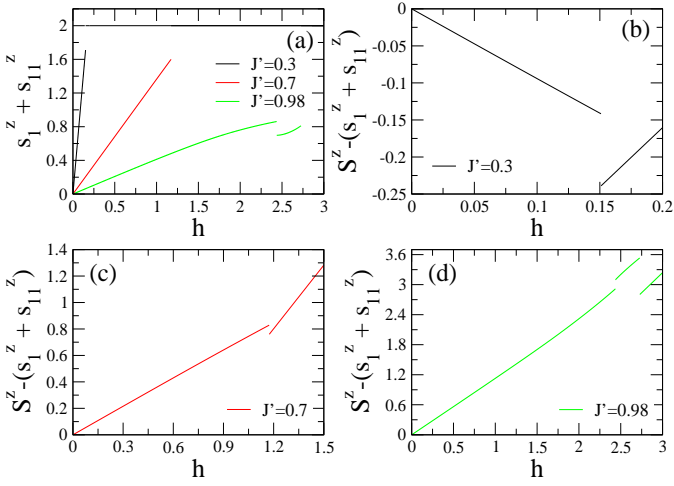}
\vspace{0pt}
\caption{Projection of (a) the net magnetization of spins 1 and 11 and (b), (c), (d) the net magnetization of the triangle strip along the magnetic field $\vec{h}$ axis. The black line corresponds to $J'=0.3$, the red to $J'=0.7$ and the green to $J'=0.98$. In (a) all lines jump to the value of 2 after their last discontinuity.
}
\label{fig:transitionfieldspins}
\end{figure}

The evolution of the discontinuity field $h_{disc}$ and the magnitude of the magnetization discontinuity with $J'$ are shown in Figs. \ref{fig:transitionfield}(a), \ref{fig:transitionfield}(c). The discontinuity evolves continuously from $J'=0$ to 1, showing that it traces back to the discontinuous magnetic response of spins 1 and 11 when $J'=0$. This can also be verified by examining its qualitative features as $J'$ is varied. When $J'$ is small (for example $J'=0.3$) spins 1 and 11 do not immediately align with the field as they now share finite exchange energy with their neighbors. They develop a component along the field direction (Fig. \ref{fig:transitionfieldspins}(a)) while characteristically the net magnetization of the other ten spins points away from the field, is weak and has increasing magnitude for small fields (Fig. \ref{fig:transitionfieldspins}(b)). The exchange energy of the triangular strip is also getting more AFM in character. This is because it is energetically favorable for the icosahedron to gain magnetic energy through the two loosely coupled spins, while the triangle strip gains exchange at the cost of magnetic energy. Thus the magnetization of the icosahedron is still dominated by the two extra spins for small magnetic fields, and in a way these screen the rest from the field.
Right after the discontinuity the two extra spins fully polarize and the triangle strip discontinuously assumes an even more negative net magnetization with a common polar angle for all spins larger than $\frac{\pi}{2}$, while the exchange energy within the triangular strip keeps getting more AFM. It reaches a minimum above the discontinuity when $h=J'$, where according to Eq. (\ref{eqn:Hamiltoniantwoparts}) the Hamiltonian is the one of an isolated triangle strip. At the same field the net magnetization of the strip is zero in order to minimize its exchange energy, and the exchange energy contribution of spins 1 and 11 is also zero. The magnetization response above the discontinuity remains the one of an isolated triangle strip as when $J'=0$, with the effective field according to Eq. (\ref{eqn:Hamiltoniantwoparts}) reduced by $J'$. Thus the saturation field is weaker from the one of the isolated triangle strip by $J'$.

Fig. \ref{fig:transitionfield}(c) shows that the strength of the discontinuity decreases up to $J'=0.44722$ where it becomes zero and the magnetization response changes qualitatively. At this value of $J'$ the derivative of $h_{disc}$ with respect to $J'$ is discontinuous, and the magnetization discontinuity changes to one of the susceptibility exactly at $h_{disc}=J'$. Below the discontinuity it is $\vec{s}_2+\vec{s}_3+\vec{s}_4+\vec{s}_5+\vec{s}_8=0$ and $\vec{s}_6+\vec{s}_7+\vec{s}_9+\vec{s}_{10}+\vec{s}_{12}=0$, and according to Eq. (\ref{eqn:Hamiltoniantwoparts}) the magnetization response is determined solely by spins 1 and 11 and has constant susceptibility like the higher field configuration. For $J' > 0.44722$ the magnetization discontinuity appears again. For example for $J'=0.7$ right after the discontinuity spins 1 and 11 again jump to their saturation magnetization (Fig. \ref{fig:transitionfieldspins}(a)), while the net magnetization of the triangle strip along the field direction is always positive and jumps to a smaller value only to become bigger with increasing field (Fig. \ref{fig:transitionfieldspins}(c)). The exchange energy of the triangle strip still gets more AFM below and exactly at the discontinuity but right above it the increasing field weakens its AFM character.

For $J' \sim 0.8$ below the discontinuity the polar angles of spins 1 and 11 start to differ, and one of the two increases with the field and can even be larger than $\frac{\pi}{2}$ and have a negative component along the field.
At $J'=0.96382$ a new discontinuity appears for lower fields, whose corresponding field $h_{disc}$ and magnetization change are shown in Figs \ref{fig:transitionfield}(b) and \ref{fig:transitionfield}(d). Its magnitude is weaker in comparison with the other discontinuity, and in contrast with it the total spin of spins 1 and 11 decreases, for example for $J'=0.98$ (Fig. \ref{fig:transitionfieldspins}(a)), while the net magnetization of the triangle strip increases (Fig. \ref{fig:transitionfieldspins}(d)). Similarly to the other discontinuity the exchange energy of the triangular strip becomes more AFM, even though its net spin increases. However for fields smaller and bigger the exchange energy of the strip becomes less AFM. This discontinuity disappears exactly at the spatially isotropic limit $J'=1$.

\section{Correspondence between Quantum and Classical Low Energy Spectrum}
\label{sec:lowenergyspectrum}

\subsection{Ground State Correlations}
\label{subsec:groundstatecorrelations}

\begin{figure}
\includegraphics[width=3.5in,height=2.8in]{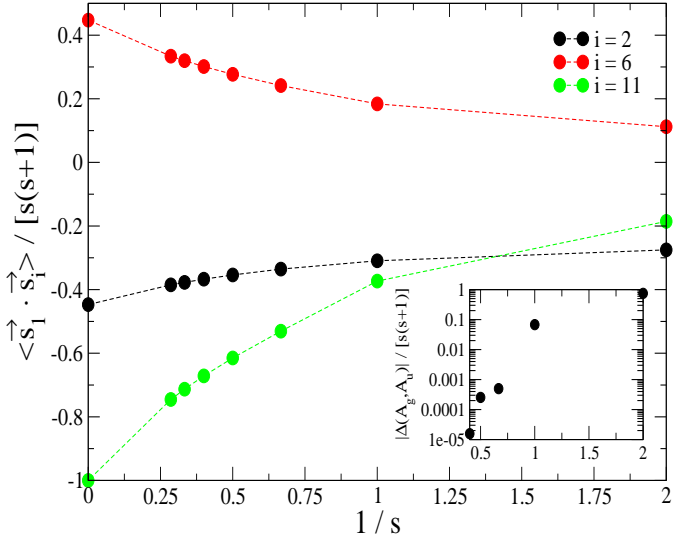}
\vspace{0pt}
\caption{Ground state distinct correlations for different $s$ as functions of $\frac{1}{s}$. The right end corresponds to the full quantum case $s=1/2$, while the left end to the classical case $s \to \infty$. The dashed lines are guides to the eye. The location of spins $\vec{s}_i$ can be seen in Fig. \ref{fig:icosahedronclusterconnectivity}. The inset shows the absolute energy difference between the lowest states in the $A_g$ and $A_u$ irreducible representations, which is also the gap to the first excited state for $s>\frac{1}{2}$.
}
\label{fig:groundstateenergycorrelation}
\end{figure}

Going back to the spatially isotropic case, the ground state of Hamiltonian (\ref{eqn:Hamiltonian}) has been calculated for $s$ ranging from $\frac{1}{2}$ to $\frac{7}{2}$ and the distinct correlations are shown in Fig. \ref{fig:groundstateenergycorrelation} as function of $1/s$. The ground state energy, given by the nearest neighbor correlation function $\vec{s}_1 \cdot \vec{s}_2$, becomes more AFM going towards the classical limit as quantum fluctuations decrease. The relative ratio of its values for $s=\frac{1}{2}$ and $\infty$ is 0.615, while for the dodecahedron the quantum fluctuations have a more drastic effect as the corresponding ratio is 0.435. The next nearest neighbor correlation $\vec{s}_1 \cdot \vec{s}_6$ acquires on the contrary a more ferromagnetic character, while the most distant correlation $\vec{s}_1 \cdot \vec{s}_{11}$ becomes even more AFM than the nearest neighbor with these two spins antiparallel in the classical limit.

A fourth-order polynomial fit of the correlation functions in $\frac{1}{s}$ reproduces their values with accuracy at least $3 \times 10^{-4}$. The formulas for the fits are:
\begin{eqnarray}
(\vec{s}_1 \cdot \vec{s}_2)_{fit} & = & -0.4472 + 0.26999 \frac{1}{s} - 0.21022 (\frac{1}{s})^2 \nonumber \\ & & + 0.097061 (\frac{1}{s})^3 - 0.018964 (\frac{1}{s})^4 \nonumber \\
(\vec{s}_1 \cdot \vec{s}_6)_{fit} & = & 0.4472 - 0.49844 \frac{1}{s} + 0.42566 (\frac{1}{s})^2 \nonumber \\ & & - 0.2508 (\frac{1}{s})^3 + 0.060347 (\frac{1}{s})^4 \nonumber \\
(\vec{s}_1 \cdot \vec{s}_{11})_{fit} & = & -1 + 1.1423 \frac{1}{s} - 1.0772 (\frac{1}{s})^2 \nonumber \\ & & + 0.76869 (\frac{1}{s})^3 - 0.20692 (\frac{1}{s})^4
\end{eqnarray}
This shows again that the magnitude of corrections in $\frac{1}{s}$ decreases fast, therefore starting from the classical ground state a few orders in perturbation theory in $\frac{1}{s}$ should give the wavefunction accurately for arbitrary $s$ \cite{NPK01}.

\subsection{Symmetric Lowest Energy Configurations for Classical Spins}
\label{subsec:classicalspin}


In the absence of a field the classical lowest energy configuration of Hamiltonian (\ref{eqn:Hamiltonian}) is a stellated geometrical structure called great icosahedron \cite{Schmidt03}, which has $I_h$ symmetry in spin space. The $SO(3)$ spin rotational symmetry of Hamiltonian (\ref{eqn:Hamiltonian}) for $h=0$ is thus reduced to the $I_h$ point symmetry group. If each spin of the icosahedron lies along one of the 12 directions of the lowest energy configuration, due to its frustrated nature there are many different ways these directions can be distributed among the spins to minimize the energy. These different ways are called colorings, and have been used to determine the minimum energy of the icosidodecahedron \cite{Axenovich01} and to analyze the lowest energy spectrum structure of the cuboctahedron and the icosidodecahedron for higher $s$ \cite{Rousochatzakis08}.
For the zero field configuration there are 120 different colorings
which all belong to the same class, meaning that any two of the colorings can be transformed to one another by application of an operation of $I_h$. Their decomposition in irreducible representations of the spatial $I_h$ group is \cite{Altmann94}
\begin{eqnarray}
& & A_g \oplus 3 T_{1g} \oplus 3 T_{2g} \oplus 4 F_g \oplus 5 H_g \oplus \nonumber \\ & & A_u \oplus 3 T_{1u} \oplus 3 T_{2u} \oplus 4 F_u \oplus 5 H_u
\label{decompositions:zerofield}
\end{eqnarray}
(the irreducible representation degeneracies are as described in Fig. \ref{fig:spectrums=5/2}). Considering the full symmetry group of the Hamiltonian, the decomposition of the 120 colorings simultaneously to the spatial $I_h$ group and the icosahedral $I$ group in spin space is further calculated, where $I$ is generated from the full $I_h$ spin space symmetry group where the lowest configuration belongs to by excluding inversion symmetry.
The characters of rotations by $\phi$ in $SO(3)$ are $\chi^{S}(\phi)=\frac{sin[(S+\frac{1}{2})\phi]}{sin\frac{\phi}{2}}$ for an irreducible representation of total spin $S$. A non-zero contribution to the characters of the total symmetry group comes from combinations of real and spin space operations \cite{Rousochatzakis08}. The decomposition of the 120 colorings for $S \leq 6$ is given in Table \ref{table:spinsymmetryzerofield}.
It respects spatial inversion symmetry with the symmetric configurations coming in pairs including both parities.

In the high field configuration two spins are aligned with the field and each of the rest points along its own direction in spin space (Sec. \ref{sec:modelandclassicallowestenergyconfiguration}). The number of different colorings that reproduce the lowest energy is again 120, and they all belong to the same class. Their decomposition in irreducible representations of the spatial $I_h$ symmetry group is also given by Eq. (\ref{decompositions:zerofield}).
Spin space symmetry can again be taken into account to further characterize the symmetrized configurations. A non-zero magnetic field breaks the $SO(3)$ symmetry in Hamiltonian (\ref{eqn:Hamiltonian}) and the symmetry operations in spin space are rotations around the field axis and vertical mirror planes that include the field axis \cite{Rousochatzakis08}.
The rotations are clockwise and counterclockwise by $\frac{\pi}{5}$, $\frac{2\pi}{5}$, $\frac{3\pi}{5}$, $\frac{4\pi}{5}$ and $\pi$, and the reflections are on five vertical mirror planes that include spins diametrically opposite in the azimuthal plane, and on five more containing the lines that bisect the difference of the azimuthal angles of two consecutive spins. The characters of the operations in spin space are given by the characters of the $C_{\infty \upsilon}$ symmetry group. The decomposition of the 120 colorings for different $S^z$ is given in Table \ref{table:spinsymmetryhighfield}. The periodicity with respect to $S^z$ is equal to 10, reflecting the symmetry structure of the spins not aligned with the field axis. The parity with respect to spatial inversion is broken and alternates with $S^z$, in contrast to what happens for fields below the discontinuity. The irreducible representations of the first five $S^z$ sectors of Table \ref{table:spinsymmetryhighfield} are exactly repeated in the next five but with opposite parity.

\subsection{Symmetric Low Energy Spectrum for $s=\frac{5}{2}$}
\label{subsec:spinfiveover2}

\begin{figure}
\includegraphics[width=3.5in,height=2.5in]{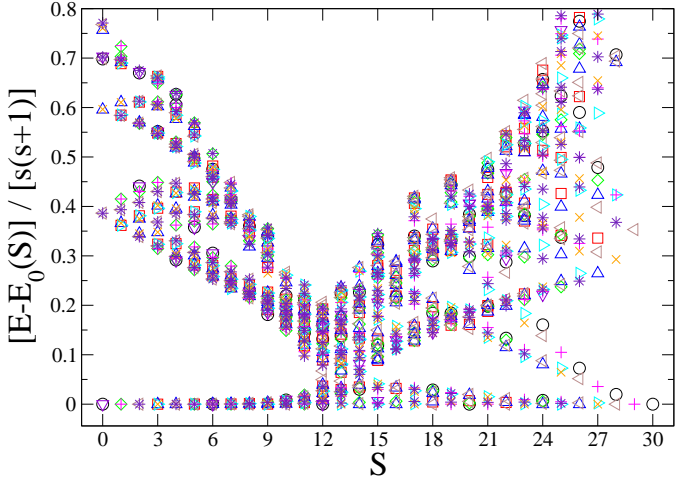}
\vspace{0pt}
\caption{Low energy spectrum for $s=\frac{5}{2}$ as function of total spin $S$ with respect to the lowest energy of each $S$ sector $E_0(S)$. Each level belongs to an irreducible representation according to the key (the number in the parenthesis is the spatial degeneracy): $\circ$: $A_g$ (1), {\color{red}{$\Box$}}: $T_{1g}$ (3), {\color{green}{$\diamond$}}: $T_{2g}$ (3), {\color{blue}{$\vartriangle$}}: $F_g$ (4), {\color{brown}{$\triangleleft$}}: $H_g$ (5), {\color{violet}{$\triangledown$}}: $A_u$ (1), {\color{cyan}{$\triangleright$}}: $T_{1u}$ (3), {\color{magenta}{+}}: $T_{2u}$ (3), {\color{orange}{$\times$}}: $F_u$ (4), {\color{indigo(dye)}{$\ast$}}: $H_u$ (5). The spatial degeneracy has to be multiplied with the $2S+1$ degeneracy of each $S$ multiplet to give the total number of states corresponding to each symbol. Only a certain number of levels has been calculated for each $S$, which explains the lack of more excited states for intermediate $S$.
}
\label{fig:spectrums=5/2}
\end{figure}

\begin{figure}
\includegraphics[width=3.5in,height=2.5in]{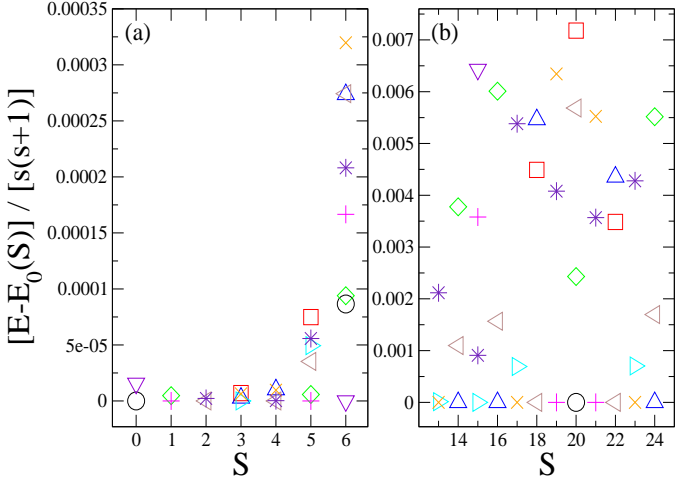}
\vspace{0pt}
\caption{Low energy spectrum close to the ground state for $s=\frac{5}{2}$ for (a) low total spin $S$ (compare with Table \ref{table:spinsymmetryzerofield}) and (b) high $S$ (compare with Table \ref{table:spinsymmetryhighfield}). The symbols are the one of Fig. \ref{fig:spectrums=5/2}.
}
\label{fig:spectrums=5/2focus}
\end{figure}

To investigate if there is a trace of the lowest energy classical configurations for low $s$, the case $s=\frac{5}{2}$ is considered first. The low energy spectrum is shown in Fig. \ref{fig:spectrums=5/2}, with each state characterized by the irreducible representation of the $I_h$ group it belongs to. It is seen that for $S$ below the discontinuity the low lying set of states is clearly separated from higher ones. This set is shown in greater detail in Fig. \ref{fig:spectrums=5/2focus}(a). Comparing with the symmetry of the classical configurations in Table \ref{table:spinsymmetryzerofield} the $s=\frac{5}{2}$ low lying states are identical in symmetry, which is a strong indication that they stem from the symmetric lowest energy classical configurations of Sec. \ref{subsec:classicalspin}. Such states which are well separated from higher excited states are associated with the Anderson tower of states or rotational band, which is the equivalent in a finite system of the breaking of SU(2) symmetry in the thermodynamic limit. Unlike the cuboctahedron \cite{Rousochatzakis08}, the lowest classical configuration of the icosahedron
is associated with only a single tower of states as all the colorings are connected with spatial symmetry operations. For $S$ above the discontinuity there are also states clearly separated from the higher ones in Fig. \ref{fig:spectrums=5/2}. The lowest lying set of states is shown in Fig. \ref{fig:spectrums=5/2focus}(b). Comparison with Table \ref{table:spinsymmetryhighfield} shows that this set must also originate in the symmetrized lowest energy classical configurations for $S$ above the discontinuity. The above comparisons show that consideration of the two different classical ground states explains the structure of the low energy spectrum for the whole $S$ range.

\begin{figure}
\includegraphics[width=3.5in,height=2.5in]{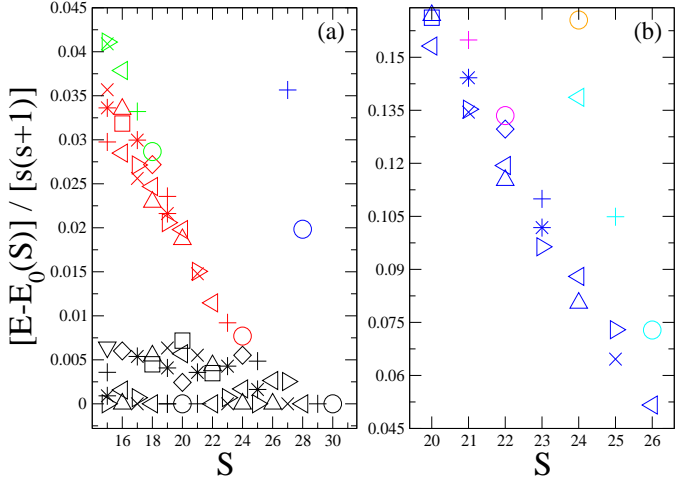}
\vspace{0pt}
\caption{Low energy spectrum for $s=\frac{5}{2}$ for high total spin $S$ broken in two parts (a) and (b). The black symbols indicate the lowest energy manifold shown in Fig. \ref{fig:spectrums=5/2focus}(b) that originates in the symmetrized lowest energy classical configirations (compare with Table \ref{table:spinsymmetryhighfield}). The groups of states with different colors have the same symmetry structure with the lowest energy manifold, therefore they correspond to classical states that share the symmetry of the lowest classical configuration but start at different values of $S$. The symbols are the one of Fig. \ref{fig:spectrums=5/2}.
}
\label{fig:spectrums=5/2focusrepetition}
\end{figure}

Looking at Fig. \ref{fig:spectrums=5/2} it is also seen that for higher $S$ there are excited states much closer in energy to the lowest lying set in comparison with what happens for small $S$. A finer plot of the low energy spectrum for higher $S$ is shown in Fig. \ref{fig:spectrums=5/2focusrepetition}. There the lowest set of states that originate in the symmetrized classical configurations is grouped together, along with sets of states where comparison shows that their symmetry structure is precisely identical to the lowest lying set. This points to the fact that they originate in excited classical configurations close to the one with minimum energy that have the same symmetry with it, but their spin is higher which explains why they start out at an $S$ different from the one corresponding to the quantum ferromagnetic configuration.

\subsection{Symmetric Low Energy Spectrum for $s < \frac{5}{2}$}
\label{subsec:spinlessthanfiveover2}

\begin{figure}
\includegraphics[width=3.5in,height=2.8in]{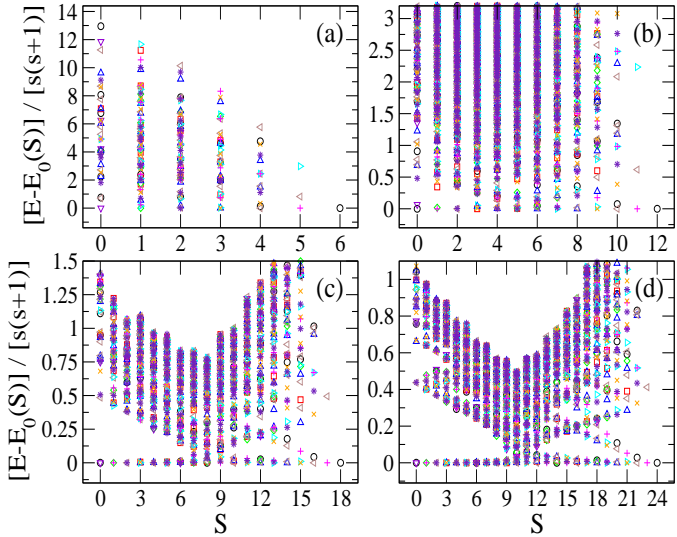}
\vspace{0pt}
\caption{Low energy spectrum as function of total spin $S$ for (a) $s=\frac{1}{2}$, (b) $s=1$, (c) $s=\frac{3}{2}$ and (d) $s=2$ with respect to the lowest energy of each $S$ sector $E_0(S)$. The symbols are the ones of Fig. \ref{fig:spectrums=5/2}.
For $s=\frac{3}{2}$ and 2 only a certain number of levels has been calculated for each $S$, which explains the lack of more excited states for intermediate $S$.
}
\label{fig:spectrums=1/2to2}
\end{figure}

\begin{figure}
\includegraphics[width=3.5in,height=2.5in]{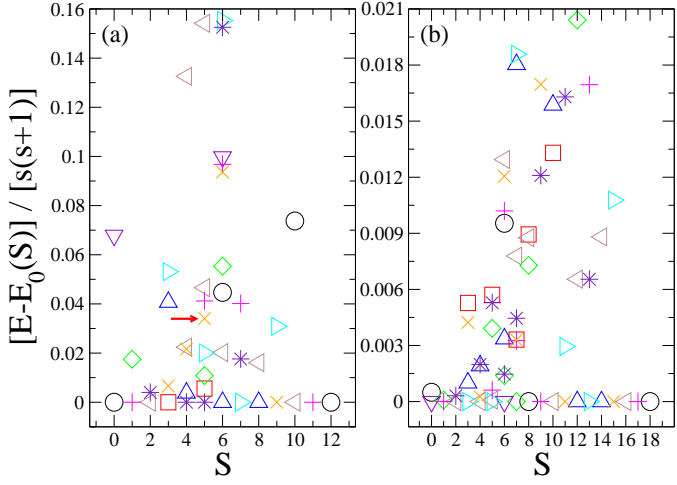}
\vspace{0pt}
\caption{Low energy spectrum close to the ground state for (a) $s=1$ and (b) $s=\frac{3}{2}$. For low total spin $S$ the results have to be compared with Table \ref{table:spinsymmetryzerofield}, while for high $S$ with Table \ref{table:spinsymmetryhighfield}. The red arrow points to the level that breaks the level ordering predicted by the symmetrized classical lowest energy manifold by having lowest energy than the two right above it. The symbols are the ones of Fig. \ref{fig:spectrums=5/2}.
}
\label{fig:spectrums=1and3/2focus}
\end{figure}

Now that the correspondence between the symmetric lowest energy classical configurations and the $s=\frac{5}{2}$ low energy spectrum is established, the next step is to examine for how low $s$ it survives. Fig. \ref{fig:spectrums=1/2to2} shows the low energy spectrum for $s$ ranging from $\frac{1}{2}$ up to 2. The energy spectrum down to $s=1$ has the same low energy structure with the one of $s=\frac{5}{2}$, with the lowest lying states clearly separated from higher excited states in the same fashion. This is shown in greater detail in Fig. \ref{fig:spectrums=1and3/2focus} for $s=1$ and $\frac{3}{2}$. The lowest lying levels for $S \leq 6$ belong to exactly the same irreducible representations as the ones for $s=\frac{5}{2}$ shown in Fig. \ref{fig:spectrums=5/2focus}(a). The only exception is the $S=5$ level pointed by the arrow for $s=1$ which has less energy than two of the levels listed in Table \ref{table:spinsymmetryzerofield}. For higher spins the comparison can be made for $s=\frac{3}{2}$ for $S=8$ up to 13, where the lowest lying levels belong to exactly the representations in Table \ref{table:spinsymmetryhighfield}. For $s=1$ the comparison can only be made and is successful for $S=7$ as there are not enough levels, but even the levels included for higher $S$ have the same symmetry in comparison with the low energy levels of higher $s$. For higher $S$ comparison of only the lowest lying level for each $S$ between $s=1$, $\frac{3}{2}$ and $\frac{5}{2}$ also shows agreement. In contrast, for low $S$ the states come in pairs which differ in spatial inversion symmetry and their order in energy in general changes with $s$. The energy difference between such states decreases with $s$ so that they become degenerate in the classical limit according to Table \ref{table:spinsymmetryzerofield}. This is exemplified by the absolute energy difference $|\Delta(A_g,A_u)|$ between the lowest lying $A_g$ and $A_u$ states that are the ground states for different $s$ and is shown in the inset of Fig. \ref{fig:groundstateenergycorrelation}. This energy difference is also the gap to the first excited state for $s > \frac{1}{2}$, and quickly decreases with $s$.

\section{Magnetization in a Field}
\label{sec:magnetizationinafield}
Since the magnetic response of Hamiltonian (\ref{eqn:Hamiltonian}) is discontinuous at the classical level, the discontinuity must show up for at least higher values of $s$ in the quantum case, and it was shown that the minimum required $s$ equals 4 \cite{Schroeder05}.
The magnetization response for a given $S$ is given by the lowest lying state.
For $S$ below the discontinuity the periodicity of the symmetry of the lowest energy state as function of $S$ is six. It belongs to the irreducible representations $A$, $T_2$, $H$, $T_1$, $H$ and $T_2$ successively (Table \ref{table:spinsymmetryzerofield}), where the inversion symmetry index has been omitted as the lowest state could have any of the two with states differing in inversion symmetry very close in energy for finite $s$ and degenerate at the classical limit as discussed in Sec. \ref{sec:lowenergyspectrum}. The periodicity of six reflects the symmetry of the classical lowest energy state on the plane perpendicular to the field, where there are six equidistant azimuthal angles (Sec. \ref{sec:modelandclassicallowestenergyconfiguration}). Above the magnetization discontinuity the corresponding periodicity is ten similarly to the whole low energy spectrum as function of $S$ (Sec. \ref{subsec:classicalspin}), with the lowest energy state highlighted in red in Table \ref{table:spinsymmetryhighfield}.

\begin{figure}
\includegraphics[width=3.5in,height=2.5in]{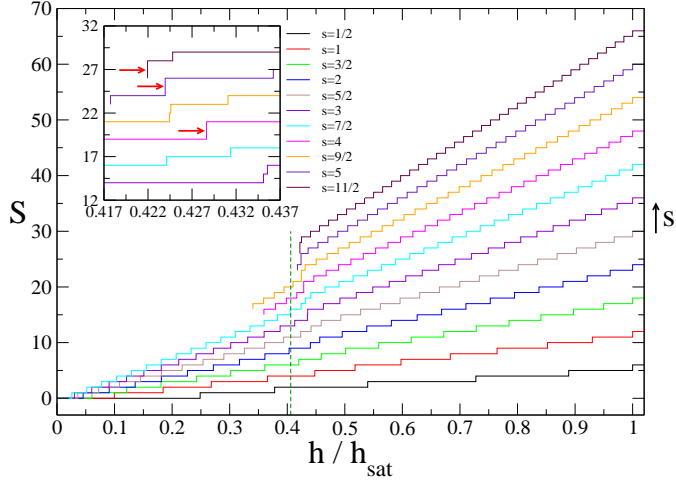}
\vspace{0pt}
\caption{Magnetization for different quantum numbers $s$ as function of the magnetic field $h$ over its saturation value $h_{sat}$. $s$ varies from bottom to top from $\frac{1}{2}$ (black line) to $\frac{11}{2}$ (maroon line) according to the legend. The dashed line shows the location of the classical discontinuity. The inset shows in detail the discontinuities for higher $s$ where $\Delta S=2$ (red arrows), and the lack of discontinuity for $s=\frac{9}{2}$.
}
\label{fig:magnetization}
\end{figure}

The magnetization for different $s$ is shown in Fig. \ref{fig:magnetization}. The calculation is extended to $s=\frac{9}{2}$, 10 and 11 for $S$ around the discontinuity with respect to Ref. \cite{Schroeder05}. When $s=\frac{9}{2}$ the discontinuity disappears, and in its place there is a spin sector that is the ground state only for a very narrow range of fields. The discontinuity reappears for $s=5$ and $\frac{11}{2}$, indicating that its disappearance for $s=\frac{9}{2}$ is only a transient effect.

\section{Conclusions}
\label{sec:conclusions}

The AHM on the icosahedron was investigated in light of its unexpected properties \cite{Schroeder05,NPK05}. First the role of its particular connectivity in the discontinuous magnetization was established with the discontinuity traced back to the magnetic response of two isolated spins brought to interact with the remaining ten spins of the cluster. Then the importance of the symmetrized lowest energy classical configurations, which also relate to the connecivity of the cluster, was revealed by exhibiting that the lowest quantum states can be traced back to them down to relatively low $s$ and are well separated from higher excited states, showing that the classical description is a good approximation for the AHM on the icosahedron. This was further corroborated by the ground state correlation functions, which can be well approximated by a small order expansion in $\frac{1}{s}$. The triangular lattice and its depleted versions that have towers of states in their low energy spectrum show that the polygon unit that makes up a cluster is important for its magnetic properties.

The results of this paper establish that magnetization discontinuities can be caused not only by a few different polygons compared with the basic unit that makes up a cluster \cite{Coffey92,NPK07} or defects in the structure in the form of a vacancy \cite{Wollny12}, but also from extra spins connected to a structure that has a smooth magnetization curve as in the icosahedron. The methods in this paper can be used to analyze magnetic properties of different small entities as function of their connectivity and establish more concrete connections between the two.

The author is very thankful to I. Rousochatzakis for discussions.

\bibliography{papericosahedron}

\begin{table}[h]
\begin{center}
\caption{Decomposition of the 120 colorings of the lowest classical configuration in zero field taking into account the spatial symmetry group $I_h$ and the spin symmetry subgroup $I$. The irreducible representation degeneracies are as described in Fig. \ref{fig:spectrums=5/2}.
There are $2(2S+1)$ symmetric configurations for each $S$ with each one $2S+1$ times degenerate, therefore the total number of states for a specific $S$ is $2(2S+1)^2$. The parity with respect to spatial inversion is not broken.}
\begin{tabular}{c|c}
 $S$ & Irreducible Representations \\
\hline
0 & $A_g \oplus A_u$ \\
\hline
1 & $T_{2g} \oplus T_{2u}$ \\
\hline
2 & $H_g \oplus H_u$ \\
\hline
3 & $T_{1g} \oplus F_g \oplus T_{1u} \oplus F_u$ \\
\hline
4 & $F_g \oplus H_g \oplus F_u \oplus H_u$ \\
\hline
5 & $T_{1g} \oplus T_{2g} \oplus H_g \oplus T_{1u} \oplus T_{2u} \oplus H_u$ \\
\hline
6 & $A_g \oplus T_{2g} \oplus F_g \oplus H_g \oplus A_u \oplus T_{2u} \oplus F_u \oplus H_u$ \\
\end{tabular}
\label{table:spinsymmetryzerofield}
\end{center}
\end{table}


\begin{table}[h]
\begin{center}
\caption{Decomposition of the 120 colorings of the lowest classical configuration for $S^z$ above the discontinuity taking into account the spatial symmetry group $I_h$ and the spin symmetry group $C_{\infty \upsilon}$. The irreducible representation degeneracies are as described in Fig. \ref{fig:spectrums=5/2}. The periodicity with respect to $S^z$ is 10, with $S^z_0$ denoting an appropriate arbitrary value of $S^z$. The parity with respect to spatial inversion is broken and alternates with $S^z$. The irreducible reprsentations of the first five $S^z$ sectors are exactly repeated in the next five but with opposite parity. The total number of states for each $S^z$ is equal to 12. The symmetric configurations highlighted in red have the lowest energy for a specific $S^z$ and finite $s$.}
\begin{tabular}{c|c}
$S^z$-$S^z_0$ & Irreducible \\
 & Representations \\
\hline
0 & ${\color{red} A_g} \oplus T_{1g} \oplus T_{2g} \oplus H_g$ \\
\hline
1 & ${\color{red} T_{2u}} \oplus F_u \oplus H_u$ \\
\hline
2 & $T_{1g} \oplus F_g \oplus {\color{red} H_g}$ \\
\hline
3 & $T_{1u} \oplus {\color{red} F_u} \oplus H_u$ \\
\hline
4 & $T_{2g} \oplus {\color{red} F_g} \oplus H_g$ \\
\hline
5 & $A_u \oplus {\color{red} T_{1u}} \oplus T_{2u} \oplus H_u$ \\
\hline
6 & $T_{2g} \oplus {\color{red} F_g} \oplus H_g$ \\
\hline
7 & $T_{1u} \oplus {\color{red} F_u} \oplus H_u$ \\
\hline
8 & $T_{1g} \oplus F_g \oplus {\color{red} H_g}$ \\
\hline
9 & ${\color{red} T_{2u}} \oplus F_u \oplus H_u$ \\
\end{tabular}
\label{table:spinsymmetryhighfield}
\end{center}
\end{table}

%
%
%
%

\end{document}